# Simulation of runaway electron generation in fusion grade tokamak and suppression by impurity injection

**Ansh Patel[1], Santosh P. Pandya[2]**

[1]*School of Liberal Studies, Pandit Deendayal Petroleum University, Gandhinagar, India*
[2]*Institute for Plasma Research, Bhat, Gandhinagar, India.*

e-mail: psantosh@ipr.res.in

## Abstract

During disruptions in fusion-grade tokamaks like ITER, large electric fields are induced following the thermal quench (TQ) period which can generate a substantial amount of Runaway Electrons (REs) that can carry up to 10 MA current with energies as high as several tens of MeV [1-3] in current quench phase (CQ). These runaway electrons can cause significant damage to the plasma-facing-components due to their localized energy deposition. To mitigate these effects, impurity injections of high-Z atoms have been proposed [1-3]. In this paper, we use a self-consistent 0D tokamak disruption model as implemented in PREDICT code [6] which has been upgraded to take into account the effect of impurity injections on RE dynamics as suggested in [4-5]. Dominant RE generation mechanisms such as the secondary avalanche mechanism as well as primary RE-generation mechanisms namely Dreicer, hot-tail, tritium decay and Compton scattering (from γ-rays emitted from activated walls) have been taken into account. These different RE-generation mechanisms provides seed REs of different amount and corresponding maximum amplitude of RE-current. In these simulations, the effect of impurities is taken into account considering collisions of REs with free and bound electrons as well as scattering from full and partially-shielded nuclear charge. These corrections were also implemented in the relativistic test particle model to simulate RE-dynamics in momentum space. We show that the presence of impurities has a non-uniform effect on the Runaway Electron Distribution function. Low energy RE lose their energy due to collisional dissipation while the high energy RE are scattered in momentum space and dissipate their energy due to higher synchrotron backreaction due to its dependence on total energy and pitch-angle. We also show that the combined effect of pitch-angle scattering induced by the collisions with impurity ions and synchrotron emission loss results in the faster dissipation of RE-energy distribution function [7]. The variation of different RE generation mechanisms during different phases of the disruption, mainly before and after impurity injections is reported.

*Key words: Runaway electrons, collisional dissipation, impurity injection, avalanche mechanism*

## Introduction:

Electrons in plasma are said to '*run-away*' when the Coulomb collisional drag force acting on them becomes smaller than the accelerating force due to an external electric field. While Runaway electrons (REs) are an interesting phenomenon, they can be very problematic for fusion-grade tokamaks like ITER, where large electric fields induced during the disruption phase can multiply a RE seed population enormously by the avalanche effect [1]. These REs can carry substantial amounts of the pre-disruption plasma current and have energies as high as few tens of MeV. The uncontrolled loss of such REs should be avoided since they deposit their energies in a highly localized manner on the plasma-facing-components and can damage them.

Massive material injection (MMI) is a possible solution to mitigate the detrimental effects of RE energy deposition [2]. Impurities such as He, Ne, Ar can be injected either in the form of solid pellets (SPI) or direct gas injection (MGI) which increases collisionality in the plasma leading to re-thermalization of low energy (~few MeV) REs and energy loss of high energy (~tens of MeV) REs.

The generation and suppression of RE during the CQ phase is the subject of this paper. The generation of runaway electrons is considered by taking into account all significant primary generation mechanisms as well as avalanching. We utilize a self-consistent calculation of electric field taking into account collisional and synchrotron drag force in the presence of impurities. The rest of the paper is structured as follows: the second section describes the model utilized for the numerical study, and the results are presented in the third section.

## Model:

A 0-D model taking into account the evolution of plasma and runaway electron (RE) current along with runway energy has been implemented in the PREDICT code [6] for disruption scenarios. The electric field is modelled taking into account replacement of ohmic current into RE current as:

$$E_{\|} = \eta(j_p - j_{RE})$$

where $j_{p,RE} = I_{P,RE}/\pi a^2 \kappa$ are the total and runaway plasma current densities and $\eta$ is the plasma resistivity. The total current $I_p$ is evolved using:

$$\frac{dI_p}{dt} = -\frac{2\pi R_0}{L} E_{\|}$$

The RE density is calculated independently through the discharge due to various generation mechanisms [5]:

$$\frac{dn_r}{dt} = \left(\frac{dn_r}{dt}\right)^{Dreicer} + \left(\frac{dn_r}{dt}\right)^{Avalanche} + \left(\frac{dn_r}{dt}\right)^{\beta-decay} + \left(\frac{dn_r}{dt}\right)^{\gamma-compton} + \left(\frac{dn_r}{dt}\right)^{Hot-tail} - \left(\frac{dn_r}{\tau_{RE}}\right)^{loss}$$

from which the runaway current is calculated as $I_{RE} = n_{RE} * e\beta c\pi r^2 \kappa$. No radial runaway losses are considered which corresponds to the most pessimistic case with regards to RE generation in disruption scenarios. However re-thermalization of REs due to energy loss is considered by using the critical energy for RE generation as a cut-off point below which the test electrons are not considered as runaways.

The relativistic test particle equations that govern RE energy dynamics in momentum space including collisional and synchrotron-radiation induced losses are given as:

$$\frac{dp_\|}{dt} = eE_\| - \frac{e^4 m_e \alpha_e(\gamma)}{4\pi\varepsilon_0^2} \gamma(Z_{coll}(\gamma) + 1 + \gamma)\frac{p_\|}{p^3} - (F_{gc} + F_{gy}\frac{q_\perp^2}{q^4})\gamma^4(\beta)^3 \frac{q_\|}{q}$$

$$\frac{dp}{dt} = eE_\| \frac{p_\|}{p} - \frac{e^4 m_e \alpha_e(\gamma)}{4\pi\varepsilon_0^2}\frac{\gamma^2}{p^2} - (F_{gc} + F_{gy}\frac{q_\perp^2}{q^4})\gamma^4(\beta)^3$$

where $\alpha_e(\gamma)$ and $Z_{coll}(\gamma)$ are correctional factors that take into account the presence of impurities as calculated in [5].

## Formation of runaway beam:

We start our numerical study at the beginning of the current quench (CQ) phase and assume a RE seed of 0.1kA generated due to incomplete thermalization of the electron energy distribution function, also known as the hot-tail mechanism, with plasma temperature = 5 eV and deuterium plasma density $n_D= 10^{20}$m$^{-3}$. REs are then generated by other primary generation mechanisms: Dreicer generation, tritium decay, and Compton scattering sources taking into account corrections [5] due to the presence of impurities. The avalanching of thermal electrons (both free and bound) into the RE region occurs due to induced high electric fields. The contribution due to all primary generation mechanisms during different phases of the current quench phase in **fig.1(a)** can be seen in **fig.2(a)**. The high electric field causes avalanching of the runaway electrons which suppresses the electric field in return. The critical energy for RE generation increases gradually with the drop in the electric field which suppresses further RE generation in the later part of the CQ phase. **Fig.1(b)** shows a temporal evolution of RE-beam energy considering different generation mechanisms.

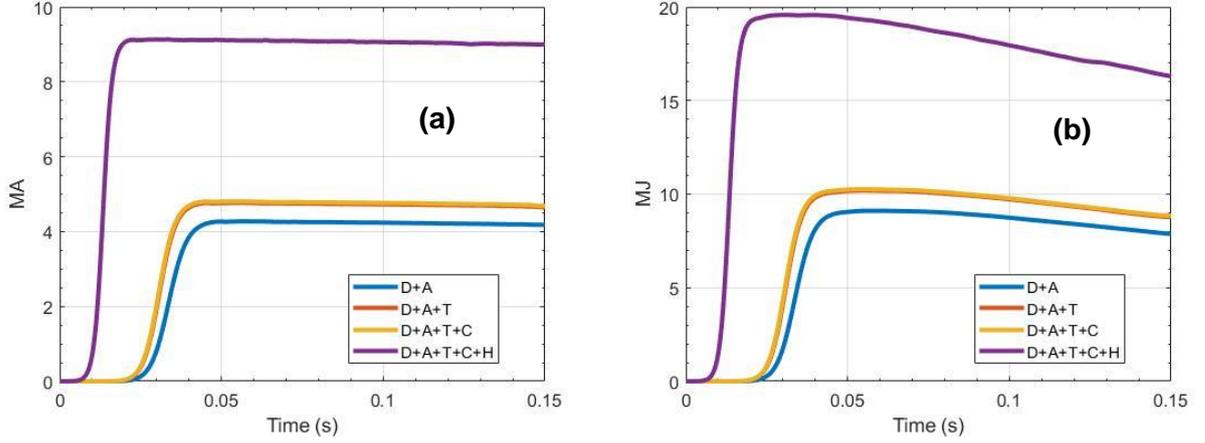

Fig. 1(a) RE generation due to various generation mechanisms. 1(b) RE beam energy due to various generation mechanisms (*D: Dreicer, A: Avalanche, T:Tritium decay, C:Compton, H: Hot tail*)

In the baseline scenario with impurity, the MMI occurs at 30 ms and we assume that the impurity Argon atoms are assimilated completely and evenly inside the plasma. The RE current and distribution function for the two baseline cases: without MMI and with $n_{Ar} = 10^{20}$ m$^{-3}$ injected at 30 ms can be seen in **fig. 2(c)** and **fig. 2(d)**.

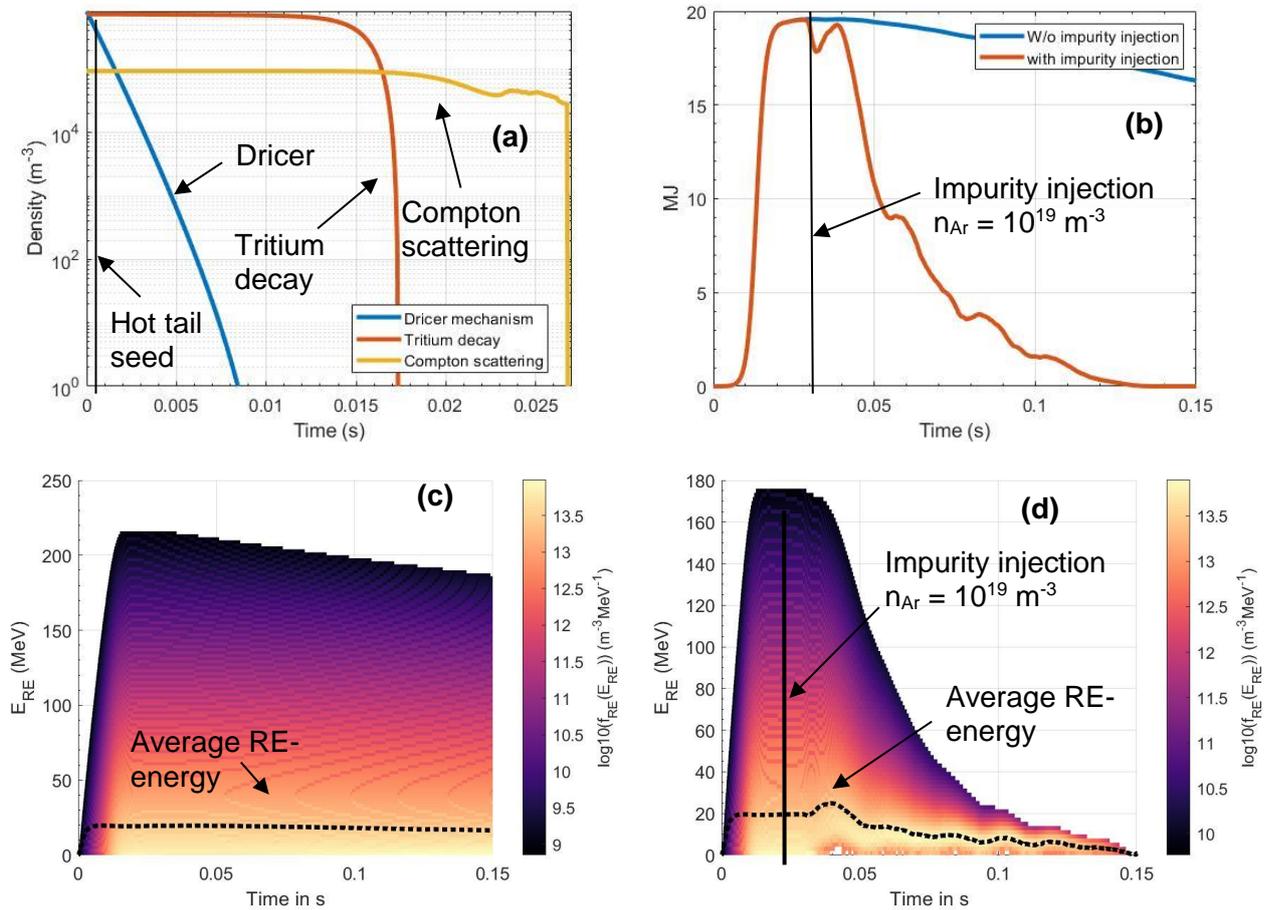

Fig. 2(a) Contribution of various generation mechanisms in the beginning of the disruption and CQ phase. 2(b) RE beam energy in absence and presence of impurity. 2(c) Runaway electron distribution function in absence of impurity. 2(d) Runaway electron distribution function in presence of impurity injection.

The amount of hot-tail seed used here is a moderate estimation. The hot-tail seed is a strong function of the thermal quench (TQ) time and pre-disruption temperature **[8]** and can carry up to a few MeV of RE current for the most pessimistic scenario. However radial losses associated with the break-up of magnetic surfaces during the thermal quench phase also have to be taken into account. Accurate estimates of the hot-tail seed would require 3D MHD simulations however it is reasonable to assume that some of the seed survives. Varying amounts of hot-tail seed due to varying discharge parameters have been shown in **fig.3 (a)** as calculated using an analytical approximation derived in **[8]** assuming an exponential drop of temperature during the thermal quench.

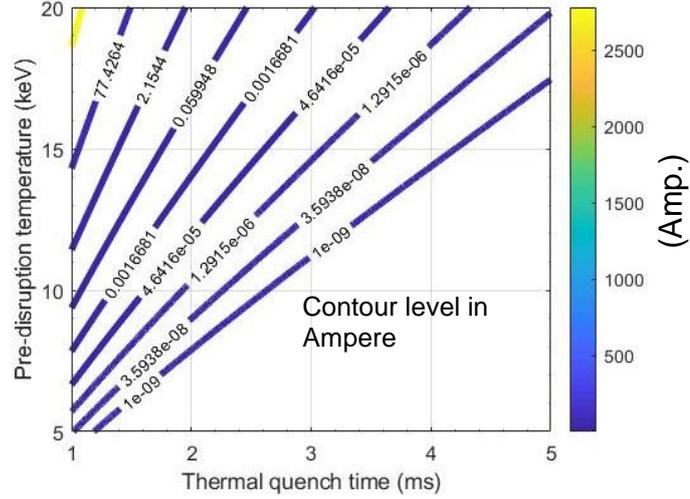

Fig. 3(a) Hot tail seed RE due to varying pre-disruption temperature and Thermal Quench time (TQ).

## Suppression of runaway beam

The RE beam starts dissipating its energy after the MMI due to collisional energy losses with the impurity ions as well as synchrotron radiation losses. The RE current also decays due to re-thermalization of the REs. A higher amount of injected impurities causes faster decay of RE current and energy due to the higher amount of energy losses as can be seen in **fig. 4(a)**. Low energy REs (~up to a few MeV) lose their energy mainly due to collisions with free, bound electrons as well as with shielded, unshielded nuclear charge. High energy REs (~tens of MeV) are pitch angle scattered in momentum space due to interactions with partially ionized impurities leading to strong synchrotron radiation losses. As shown in **fig. 4(b)**, the perpendicular momentum of REs increases drastically on impurity injections which enhances synchrotron radiation losses and consequently thermalizes the REs. In **fig.4 (c)**, the ratio of synchrotron backreaction force to the collisional drag force is shown for two different RE fractions born at different times during the discharge. For the early-born, high energy fraction (blue), the increase in perpendicular momentum enhances synchrotron losses significantly as compared to collisional losses. In contrast, for the late-born, low-energy fraction (red), collisional and synchrotron losses play almost an equal role. Hence, the presence of impurities has a two-fold effect on RE energy dissipation: the higher number of collisions decrease the RE energy and pitch-angle scattering of REs in presence of impurities also enhances synchrotron losses, especially for high energy REs.

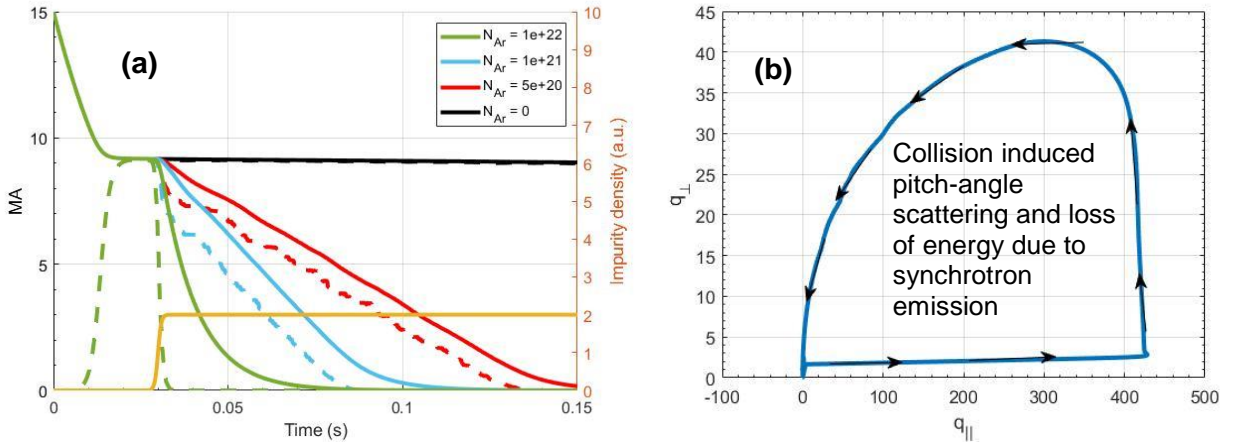

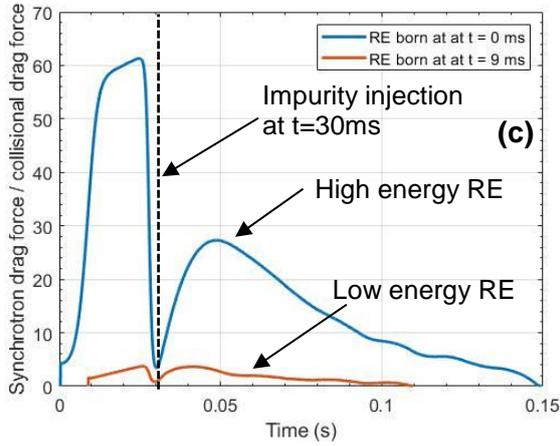 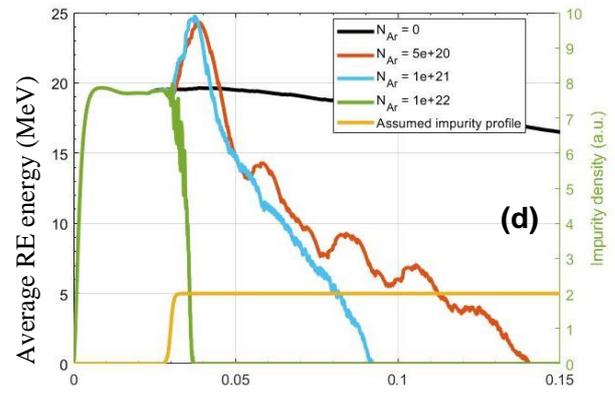

Fig.4(a) Effect of impurity amount on RE current. 4(b) 2D momentum space plot of RE showing pitch-angle scattering. 4(c) Ratio of synchrotron and collisional drag force for two different REs. 4(d) Effect of impurity amount on RE average energy

**Conclusions:**

A numerical study of RE dynamics in the presence of impurities is performed using a 0D disruption model considering significant sources RE sources that would be present in a fusion grade tokamak. The contribution of primary sources is shown to be considerable in the initial part of the CQ phase after which fast avalanching of RE current suppresses the electric field and consequently primary generation mechanisms. The presence of impurities is shown to cause decay of RE current and energy with a higher amount of impurities leading to faster decay of both parameters (RE-current and beam energy). Pitch angle scattering caused by collisions with impurity ions enhance synchrotron emissions drastically and is very significant especially for high energy REs.